\documentclass[twocolumn,superscriptaddress,aps,preprintnumbers,amsmath,amssymb,pral]{revtex4-1}

\usepackage{graphicx}
\usepackage{epstopdf}
\usepackage{dcolumn}
\usepackage{bm}
\usepackage{hyperref}
\usepackage{color}
\usepackage{cancel}

\begin{document}


\def\a{\alpha}
\def\b{\beta}
\def\c{\varepsilon}
\def\d{\delta}
\def\e{\epsilon}
\def\f{\phi}
\def\g{\gamma}
\def\h{\theta}
\def\k{\kappa}
\def\l{\lambda}
\def\m{\mu}
\def\n{\nu}
\def\p{\psi}
\def\q{\partial}
\def\r{\rho}
\def\s{\sigma}
\def\t{\tau}
\def\u{\upsilon}
\def\v{\varphi}
\def\w{\omega}
\def\x{\xi}
\def\y{\eta}
\def\z{\zeta}
\def\D{\Delta}
\def\G{\Gamma}
\def\H{\Theta}
\def\L{\Lambda}
\def\F{\Phi}
\def\P{\Psi}
\def\S{\Sigma}

\def\o{\over}
\def\beq{\begin{eqnarray}}
\def\eeq{\end{eqnarray}}
\newcommand{\gsim}{ \mathop{}_{\textstyle \sim}^{\textstyle >} }
\newcommand{\lsim}{ \mathop{}_{\textstyle \sim}^{\textstyle <} }
\newcommand{\vev}[1]{ \left\langle {#1} \right\rangle }
\newcommand{\bra}[1]{ \langle {#1} | }
\newcommand{\ket}[1]{ | {#1} \rangle }
\newcommand{\EV}{ {\rm eV} }
\newcommand{\KEV}{ {\rm keV} }
\newcommand{\MEV}{ {\rm MeV} }
\newcommand{\GEV}{ {\rm GeV} }
\newcommand{\TEV}{ {\rm TeV} }
\newcommand{\1}{\mbox{1}\hspace{-0.25em}\mbox{l}}
\newcommand{\headline}[1]{\noindent{\bf #1}}
\def\diag{\mathop{\rm diag}\nolimits}
\def\Spin{\mathop{\rm Spin}}
\def\SO{\mathop{\rm SO}}
\def\O{\mathop{\rm O}}
\def\SU{\mathop{\rm SU}}
\def\U{\mathop{\rm U}}
\def\Sp{\mathop{\rm Sp}}
\def\SL{\mathop{\rm SL}}
\def\tr{\mathop{\rm tr}}
\def\mpl{M_{PL}}

\def\IJMP{Int.~J.~Mod.~Phys. }
\def\MPL{Mod.~Phys.~Lett. }
\def\NP{Nucl.~Phys. }
\def\PL{Phys.~Lett. }
\def\PR{Phys.~Rev. }
\def\PRL{Phys.~Rev.~Lett. }
\def\PTP{Prog.~Theor.~Phys. }
\def\ZP{Z.~Phys. }

\def\dd{\mathrm{d}}
\def\ff{\mathrm{f}}
\def\BH{{\rm BH}}
\def\inf{{\rm inf}}
\def\ev{{\rm evap}}
\def\eq{{\rm eq}}
\def\SM{{\rm sm}}
\def\Mpl{M_{\rm Pl}}
\def\GeV{{\rm GeV}}
\newcommand{\Red}[1]{\textcolor{red}{#1}}


\title{A ``Gauged" $U(1)$ Peccei-Quinn Symmetry}

\author{Hajime Fukuda}
\affiliation{Kavli IPMU (WPI), UTIAS, University of Tokyo, Kashiwa, Chiba 277-8583, Japan}
\author{Masahiro Ibe}
\affiliation{ICRR, University of Tokyo, Kashiwa, Chiba 277-8582, Japan}
\affiliation{Kavli IPMU (WPI), UTIAS, University of Tokyo, Kashiwa, Chiba 277-8583, Japan}
\author{Motoo Suzuki}
\affiliation{ICRR, University of Tokyo, Kashiwa, Chiba 277-8582, Japan}
\affiliation{Kavli IPMU (WPI), UTIAS, University of Tokyo, Kashiwa, Chiba 277-8583, Japan}
\author{Tsutomu T. Yanagida}
\affiliation{Kavli IPMU (WPI), UTIAS, University of Tokyo, Kashiwa, Chiba 277-8583, Japan}
\begin{abstract}
The Peccei-Quinn (PQ) solution to the strong $CP$ problem requires an anomalous global $U(1)$ symmetry, 
the PQ symmetry.
The origin of such a convenient global symmetry is quite puzzling from the theoretical point of view in many aspects.
In this paper, we propose a simple prescription which provides 
an origin of the PQ symmetry.
There, the global $U(1)$ PQ symmetry is virtually embedded in a gauged $U(1)$ PQ symmetry.
Due to its simplicity, this mechanism can be implemented in many conventional models with the PQ symmetry.
\end{abstract}

\date{\today}
\maketitle
\preprint{IPMU 17-0040}
\section{Introduction} \vspace{-.3cm}
The Peccei-Quinn (PQ) mechanism\,\cite{Peccei:1977hh,Peccei:1977ur,Weinberg:1977ma,Wilczek:1977pj} 
is the most successful solution to the strong $CP$ problem.
There, a  global $U(1)$ symmetry (the PQ symmetry) which is almost  exact  but broken by the axial anomaly of QCD
plays a crucial role.
After spontaneous breaking, the effective $\theta$-angle of QCD is cancelled  by the vacuum expectation value (VEV) 
of the associated pseudo Nambu-Goldstone boson, the axion $a$.

The origin of such a convenient global symmetry is, however, quite puzzling from the theoretical point of view in many aspects.
By definition, the PQ symmetry is not an exact symmetry. 
Besides, the postulation of global symmetries is not comfortable in the sense of general relativity. 
It is also argued that all global symmetries are broken by quantum gravity 
effects~\cite{Hawking:1987mz,Lavrelashvili:1987jg,Giddings:1988cx,Coleman:1988tj,Gilbert:1989nq,Banks:2010zn}.

In this paper, we address a question in which circumstances a theory admits the global PQ symmetry.
If we could regard the PQ symmetry as a $U(1)$ gauge symmetry, 
there would be no suspicion about the exactness and the consistency with  quantum gravity.
The PQ symmetry is, however, broken by the QCD anomaly, and hence, it cannot be a consistent gauge symmetry as it is. 

To circumvent the dilemma, let us recall that, for example, the $U(1)_Y$ gauge symmetry of the Standard Model would be anomalous 
if it coupled only to the lepton sector.
The anomalies of the $U(1)_Y$ gauge symmetry in the lepton sector are cancelled only when it also couples to the quark sector.
In a similar manner, it seems conceivable that the anomalies of the gauged PQ symmetry, $U(1)_{gPQ}$,
are cancelled between the contributions from two (or more) PQ charged sectors. 

To make one step forward, let us assume that the PQ charged sectors are completely decoupled with each other except for gauge
interactions.
In this limit, an additional accidental $U(1)$ symmetry appears, 
whose charge assignment coincides with the $U(1)_{gPQ}$ symmetry in each sector up to relative normalizations.
There, the accidental symmetry is broken only by the QCD anomaly, and hence, it plays the role of the global PQ symmetry
for the PQ mechanism.

The interactions between the PQ charged sectors inevitably break the accidental symmetry.
Thus, the original question about the plausibility of the global PQ symmetry
is reduced to the question how well such cross-sector symmetry breaking operators are suppressed.
To this question, the gauged PQ symmetry again provides an answer. 
The  cross-sector symmetry breaking operators can be suppressed by an appropriate charge assignment of $U(1)_{gPQ}$.
Therefore, the origin of the anomalous global PQ symmetry can be attributed to a gauged $U(1)$ PQ symmetry.

In the literature, there have been many attempts to achieve the PQ symmetry 
as an accidental symmetry resulting from (discrete) gauge 
symmetries~\cite{Kim:1981bb,Georgi:1981pu,Dimopoulos:1982my,Frampton:1981qu,Kang:1982bx,Lazarides:1982tw,Barr:1992qq,Kamionkowski:1992mf,Holman:1992us,Dine:1992vx,Dias:2002gg,
Carpenter:2009zs,Harigaya:2013vja,Harigaya:2015soa,Redi:2016esr}%
~\footnote{Discrete gauge symmetries are immune to quantum gravity effects~\cite{Krauss:1988zc,Preskill:1990bm,Preskill:1991kd,Banks:1991xj}. }.
There have also been arguments of the origin of the axion in string theory~\cite{Witten:1984dg,Kallosh:1995hi,Svrcek:2006yi} and in extra dimensional 
setups~\cite{Cheng:2001ys,Izawa:2002qk,Hill:2002kq,Fukunaga:2003sz,Izawa:2004bi,Choi:2003wr,Grzadkowski:2007xm}.

In this context, our prescription adds a simple field theoretical explanation of the origin of the PQ symmetry.
There, the PQ symmetry is virtually embedded in a gauged $U(1)$ PQ symmetry%
~\footnote{A model discussed in \cite{Cheng:2001ys} also achieve an virtual embedment of the PQ symmetry 
in a gauged $U(1)$ PQ symmetry in an extra dimensional setup. }.
Due to its simplicity, this mechanism can be implemented in many conventional  models with the PQ symmetry.
We also emphatically refer \cite{Lazarides:1982tw,Barr:1982uj,Choi:1985iv} which discuss the domain wall problems of axion models with similar structures 
we consider in the following.

\vspace{-.4cm}

\section{General prescription} \vspace{-.3cm}
Let us recall invisible axion models such as the KSVZ model~\cite{Kim:1979if,Shifman:1979if} or the DSFZ model~\cite{Dine:1981rt,Zhitnitsky:1980tq}.
There, the postulated anomalous global PQ symmetry is spontaneously broken with which the axion field associates.  
The non-perturbative effects of QCD generate the axion potential through the axial anomalies.

Now let us bring two sectors of the invisible axion models.
The two PQ symmetries in each sector, $U(1)_{PQ}$ and $U(1)_{PQ'}$,  are explicitly broken by the QCD anomalies, and the corresponding Noether currents 
$j_{PQ}^\mu$ and $j_{PQ'}^\mu$ satisfy the anomalous ward identities,
\begin{eqnarray}
\label{eq:anom}
\q j_{PQ} = \frac{g_s^2}{32\pi^2}N_1 G \tilde G\ ,\quad
\q j_{PQ'} = \frac{g_s^2}{32\pi^2}N_2 G \tilde G\ .
\end{eqnarray}
Here, $G$ the gauge field strength of QCD, $g_s$ the QCD coupling constant. 
The Lorentz indices and the color indices are suppressed. 
The coefficients $N_1$ and $N_2$ depend on each invisible axion model.

In the two anomalous symmetries, there is a linear combination which is free from the QCD anomaly.
Hereafter, we consider that the anomaly free combination is a gauge symmetry, which we name the $U(1)_{gPQ}$ symmetry.
Here, we assume that the $U(1)_{gPQ}$ is free from all anomalies%
~\footnote{The $[U(1)_{gPQ}]^3$ anomaly and the gravitational anomaly of $U(1)_{gPQ}$ can be cancelled by adding fermions
which are singlet under the Standard Model gauge groups.}.

In each sector, breaking operators of the global PQ symmetries are forbidden by the 
$U(1)_{gPQ}$ symmetry.
Therefore, the $U(1)_{gPQ}$ symmetry provides protection of the PQ symmetries in each sector.

Let us further assume that there are no interactions between the two sectors except for the gauge interactions.
In this limit, the PQ symmetries in each sector are broken only by the anomalies.
It should be noted  that the radiative corrections generate interactions between the two sectors. 
Those corrections, however, do not break the PQ symmetries in each sector 
since they are broken only by the $U(1)_{gPQ}$ and the QCD anomalies.
Therefore, in this limit, the theory possesses an accidental $U(1)$ symmetry in addition to the $U(1)_{gPQ}$ gauge symmetry.
In the following, we call this anomalous accidental symmetry, $U(1)_{aPQ}$.
As it has been noted, the $U(1)_{aPQ}$ symmetry plays the role of the PQ symmetry for the PQ mechanism.

In reality,  there are interaction terms between the two sectors.
In particular, there are  terms which are invariant under the $U(1)_{gPQ}$ gauge symmetry but 
break the $U(1)_{aPQ}$ symmetry. 
For example, let us consider operators ${\cal O}_1$ and ${\cal O}_2$
which consist of fields in each sector, respectively.
When these two operators have non-vanishing and opposite $U(1)_{gPQ}$ charges,
the interaction terms 
\begin{eqnarray}
\label{eq:explicit}
{\cal L}_{\cancel{aPQ}} = \frac{1}{ 
M_{PL}^{
d_{ {\cal O}_1} 
+
d_{ {\cal O}_2} -4
} 
} {\cal O}_1 {\cal O}_2 + h.c. \ , 
\end{eqnarray}
explicitly break the $U(1)_{aPQ}$  symmetry.
Here, $d_{ {\cal O}_{1,2}}$ denote the mass dimensions of the corresponding operators,
and  $M_{\rm PL}$ denotes the reduced Planck scale.
Given the general discussion that all global symmetries are broken by quantum gravity effects, 
there is no principle to suppress these terms since it is consistent with  gauge symmetries.

Such explicit breaking terms of the $U(1)_{aPQ}$ symmetry are, however, acceptable as long as 
the breaking effects are small enough not to spoil the PQ mechanism.
In practice,  the current experimental upper limit on the $\theta$ angle, $\theta \lesssim 10^{-10}$~\cite{Baker:2006ts},
can be satisfied for $d_{ {\cal O}_1} + d_{ {\cal O}_2}  > 10$
when the PQ symmetries are spontaneously broken at $10^{10-12}$\,GeV~\cite{Barr:1992qq,Kamionkowski:1992mf,Holman:1992us}.

The mass dimensions of the lowest dimensional symmetry breaking operator depends on the charge 
assignment of $U(1)_{gPQ}$.
In fact, as we exemplify later, there are many possible charge assignments 
which suppress the $U(1)_{aPQ}$ breaking effects down to an acceptable level.

\vspace{-.4cm}

\section{Decomposition of $U(1)_{gPQ}$ and $U(1)_{aPQ}$} \vspace{-.3cm}
Before moving to explicit examples, let us discuss how to decompose the $U(1)_{gPQ}$ and the $U(1)_{aPQ}$
symmetries.
For that purpose, let us consider a simple example where the invisible axion candidates in the two sectors
correspond to the axial components of complex SM gauge singlet scalar fields $\phi$ and $\phi'$,
\begin{eqnarray}
\label{eq:axialcomp}
\phi = \frac{1}{\sqrt{2}} f_a\, e^{i \tilde a/f_a}\ , \quad 
\phi' = \frac{1}{\sqrt{2}}f_b\, e^{i \tilde b/f_b}\ .
\end{eqnarray}
Here, $f_{a,b}$ are the decay constants of each sector and we keep only the axial components, $\tilde a$ and $\tilde b$.
The domains of them are given 
\begin{eqnarray}
\label{eq:domain}
\tilde{a}/f_a = [0, 2\pi)\ ,  \quad \tilde{b}/f_b = [0, 2\pi)\ ,  
\end{eqnarray}
respectively.

Let us assume that the $U(1)_{gPQ}$ gauge charges of the complex scalars are $q$ and $q'$, respectively.
In this case, the axial components are shifted by,
\begin{eqnarray}
\tilde a/f_a \to{\tilde a}/{f_a} + q \a \ ,
\quad {\tilde b}/{f_b} \to {\tilde b}/{f_b} + q' \a\ ,
\end{eqnarray}
under the $U(1)_{gPQ}$ symmetry.
Hereafter, we take the normalization of $\alpha$ such that $q$ and $q'$ are relatively prime integers
without loosing generality.

From the covariant kinetic terms of $\phi$ and $\phi'$, we obtain
\begin{eqnarray}
{\cal L} &=& |D_\mu \phi|^2 + |D_\mu \phi '|^2  
\cr 
&=& \frac{1} {2} (\partial \tilde a)^2 
+ \frac{1} {2} (\partial \tilde  b)^2 
- g A_\mu (q f_a  \q^\mu \tilde a + q' f_b  \q^\mu\tilde b) \cr
&&+ \frac{g^2}{2} (q^2 f_a^2 + q'^2f_b^2 ) A_\mu A^\mu \cr
&=& \frac{1}{2}(\q  a)^2 + \frac{1}{2}m_A^2 \left(A_\mu - \frac{1}{m_A}\q_\mu  b \right)^2\ .
\end{eqnarray}
where, $g$ is the gauge coupling constant of $U(1)_{gPQ}$. 
The mass of the $U(1)_{gPQ}$ gauge boson, $A_\mu$, is given by,
\begin{eqnarray}
m_A^2 = g^2(q^2 f_a^2 + q'^2f_b^2 ) \ .
\end{eqnarray}
In the final expression, we redefine the axial fields by
\begin{eqnarray}
\left(
\begin{array}{cc}
 a   \\
 b
\end{array}
\label{eq:decomp}
\right)=
\frac{1}{\sqrt{q^2 f_a^2 + q'^2f_b^2 }}\left(
\begin{array}{cc}
q' f_b   &  -q f_a   \\
q f_a  &   q' f_b
\end{array}
\right)
\left(
\begin{array}{cc}
 \tilde a   \\
\tilde b
\end{array}
\right)\ .
\end{eqnarray}
The  field $ b$ is the would-be  Nambu-Goldstone boson, while the gauge 
invariant field $a$ corresponds to the PQ axion.

\begin{figure}[t]
\begin{center}
  \includegraphics[width=.7\linewidth]{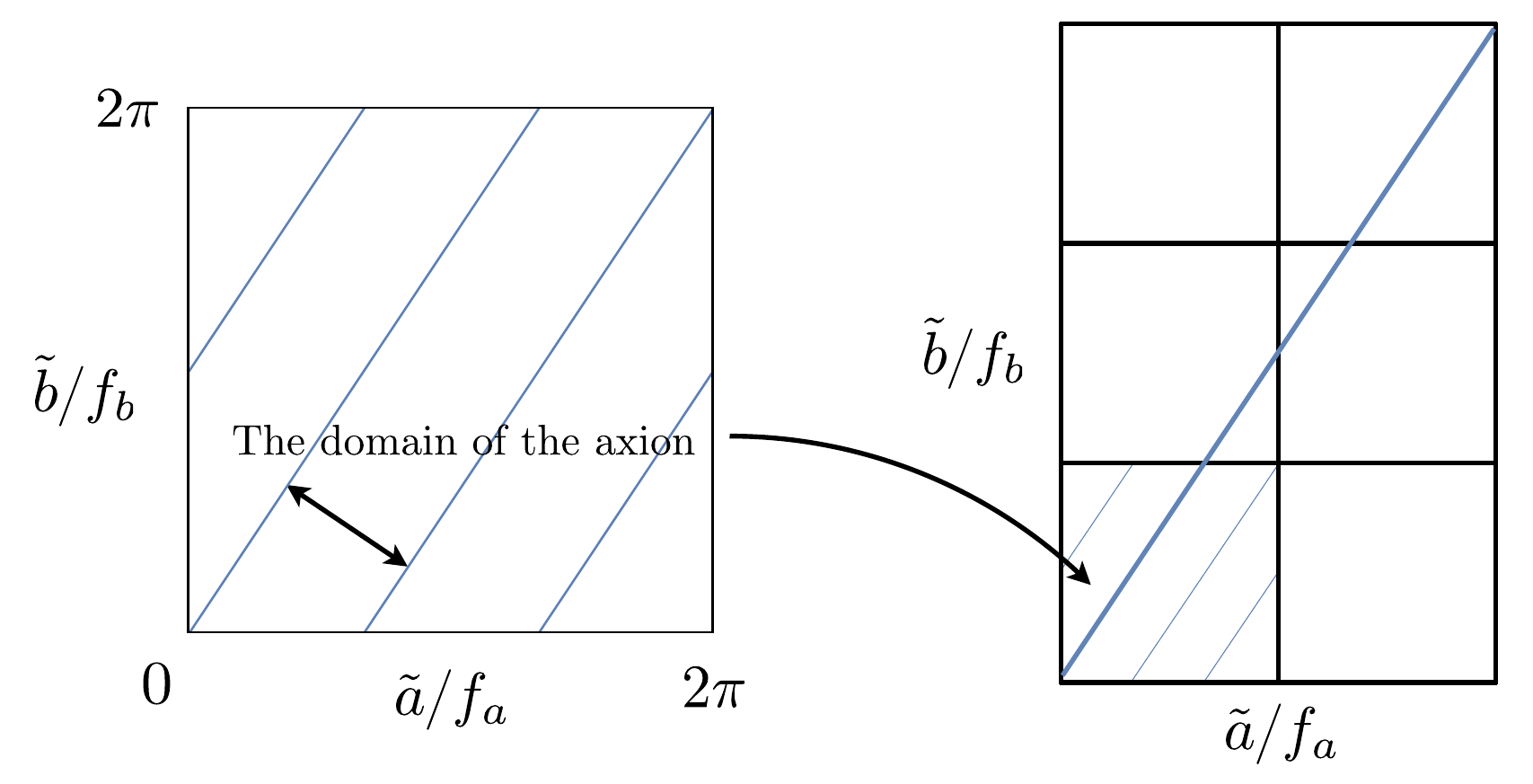}
 \end{center}
\caption{\sl \small
(Left) A gauge orbit in the domain of $(\tilde a/f_a, \tilde b/f_b)$ for $q=2$, $q' =3$.
The domain of $a$ is given by the interval between the orbits.
(Right) The unwind gauge orbits. 
}
\label{fig:domain}
\end{figure}

To extract an gauge invariant $U(1)_{aPQ}$ global symmetry, let us remember that 
a gauge orbit of $U(1)_{gPQ}$ winds the domain of $(\tilde{a}/f_a, \tilde{b}/f_b)$  more than once for $q \neq q'$
(see Fig.\,\ref{fig:domain}).
Then, the domain of $a$ is given by the interval of the gauge orbit in the domain since the field points connected by a gauge orbit
is physically equivalent.
When we take that $q$ and $q'$ are relatively prime integers, we find the axion interval in the figure is given by,
\begin{eqnarray}
\label{eq:interval}
a = \left[ 0,  \frac{2\pi f_a f_b}{\sqrt{q^2 f_a^2 + q'^2f_b^2 }} \right) \ .
\end{eqnarray}
Thus, with a decay constant,
\begin{eqnarray}
\label{eq:FA}
F_a  = \frac{f_a f_b}{\sqrt{q^2 f_a^2 + q'^2f_b^2 }}\ ,
\end{eqnarray}
the $U(1)_{aPQ}$ symmetry is realized by the shift of the axion,
\begin{eqnarray}
\label{eq:aPQ}
\frac{a}{F_a} \to \frac{a}{F_a} + \delta_{PQ} \ ,
\end{eqnarray}
with $\delta_{PQ}$ ranging from $0$ to $2\pi$%
~\footnote{We may extend our analysis where there is a kinetic mixing between $\tilde a$ and $\tilde b$,
although the kinetic mixing does not change our discussion.}.

The anomalous coupling of the axial components depends 
on models of the invisible axion models.
In order for the $U(1)_{gPQ}$ symmetry is free from the anomaly,
the anomalous coupling should appear in the form of
\begin{eqnarray}
{\cal L}_{\rm QCD} &=& \frac{g_s^2}{32\pi^2} N
\left(\frac{q'\tilde a}{ f_a} - \frac{q\tilde b}{ f_b}\right) G\tilde{G} \ , \\
&=& \frac{g_s^2}{32\pi^2} N
\frac{a}{F_a} G\tilde{G} \  .
\end{eqnarray}
Here, $N$ is a model dependent integer.


\vspace{-0.4cm}

\section{Examples}
\vspace{-.3cm}
\subsection{Barr-Seckel Model}
\label{sec:BS}
\vspace{-.3cm}
As the simplest example, let us discuss a model 
based on two KSVZ axion models~\cite{Kim:1979if,Shifman:1979if}.
This example corresponds to the model discussed in \cite{Barr:1992qq}.

In each KSVZ sector, the PQ symmetry is spontaneously broken 
by the VEVs of complex scalars $\phi$ and $\phi'$ whose PQ charges are unity.
In each sector, the scalars couple to extra vector-like quarks  via
\begin{eqnarray}
{\cal L} = y \phi Q\bar{Q} + h.c. \ ,
\end{eqnarray}
and 
\begin{eqnarray}
{\cal L} = y' \phi' Q'\bar{Q'} + h.c. \ .
\end{eqnarray}
The PQ charges of the extra quarks are taken to be $Q(0)$ and $\bar{Q}(-1)$ in the first KSVZ sector
and $Q'(0)$ and $\bar{Q}'(-1)$ in the second sector.
We assume that there are $N_f$ and $N_f'$ flavors of the extra quarks in each sector.

Due to the QCD anomaly, the axion candidates in each sector have anomalous coupling,
\begin{eqnarray}
{\cal L} = \frac{g_s^2}{32\pi^2} 
\left(\frac{N_f\,\tilde a}{ f_a} + \frac{N_f'\,\tilde b}{ f_b}\right)
G\tilde{G} \ .
\end{eqnarray}
Here, we define the axial components of the KSVZ scalars as in Eq.\,(\ref{eq:axialcomp}).
From this expression, we find that a linear combination of the two PQ symmetries
with the charge assignments $\phi(q)$ and $\phi(q')$ is  free from the anomaly for
\begin{eqnarray}
q'/q = - N_f/N_f' \ .
\end{eqnarray}
As discussed in the previous section, we regard the anomaly free PQ symmetry 
as the $U(1)_{gPQ}$ gauge symmetry,
where $q$ and $q'$ are normalized so that they are relatively prime integers.

Under the  $U(1)_{gPQ}$ symmetry,  no explicit PQ breaking operators appear in each sector.
The interaction terms between the two KSVZ sectors, on the other hand, generically break $U(1)_{aPQ}$.
In fact, the lowest dimensional operator which breaks the $U(1)_{aPQ}$ symmetry
is given by,
\begin{eqnarray}
\label{eq:explicit2}
{\cal L}_{\cancel {aPQ}} = \frac{1}{M_{\rm PL}^{|q|+|q'|-4}} \phi^{|q'|} \phi^{\prime |q|} + h.c.
\end{eqnarray}
As we have seen in the previous section, the explicit breaking of the PQ symmetry is acceptable 
when $|q|+|q'| > 10$.
Once this condition is satisfied, the anomalous $U(1)_{aPQ}$ of an acceptable quality 
appears as a result of the $U(1)_{gPQ}$ gauge symmetry.

Let us comment here that  $q$ and $q'$ in our normalization are given by,
\begin{eqnarray}
q = N_f'/n_{gcd}\ , \quad q' = - N_f/n_{gcd} \ ,
\end{eqnarray}
when $N_f$ and $N_f'$ has common divisors, $n_{gcd} > 1$. 
In this case, the anomalous coupling of the axion is given by,
\begin{eqnarray}
\label{eq:QCD2}
{\cal L}_{\rm QCD} = \frac{g_s^2}{32\pi^2} n_{gcd}
\frac{a}{F_a} G\tilde{G} \ ,
\end{eqnarray}
which means $N=n_{gcd}$ in Eq.\,(13).

\vspace{-.4cm}

\subsection{Composite Axion Model}\vspace{-.3cm}
As a second example, let us apply our prescription to the so-called composite axion model~\cite{Kim:1984pt,Choi:1985cb}
~\footnote{For other attempts to obtain a high-quality PQ symmetry in the composite axion model,
see e.g. \cite{Randall:1992ut,Redi:2016esr}.}.
There, we consider an $SU(N_c)$  gauge theory with vector-like fermions of $SU(N_c)\times$QCD quantum numbers,
\begin{eqnarray}
Q(N_c,3), \quad \bar{Q}(\bar{N_c},\bar{3}), \quad q(N_c,1), \quad \bar{q}(\bar{N_c},1)\ .
\end{eqnarray}
This model possesses an axial $U(1)$ symmetry with the charge assignments,
\begin{eqnarray}
\label{eq:SUNaxial}
Q(1), \quad \bar{Q}(1), \quad q(-3), \quad \bar{q}(-3)\ .
\end{eqnarray}
This symmetry is free from the anomaly of $SU(N_c)$ but broken by the QCD anomaly.
We identify this symmetry with the anomalous PQ symmetry in the first sector.
The anomalous PQ symmetry is spontaneously broken at the dynamical scale of $SU(N_c)$, 
where the axion appears as an composite field%
~\footnote{There are $15$ light pseudo-goldstone modes associated with the chiral symmetry breaking at $\Lambda_{N}$,
which are color charged except for the axion candidate.
The colored pseudo Nambu-Goldstone bosons obtain masses of ${\cal O}(\alpha_s \Lambda_N)$ where $\alpha_s = g_s^2/4\pi$.
See e.g.~\cite{Izawa:2002qk}. 
}.

According to the general prescription, we further introduce another sector of the composite composite axion
where $N_c$ is replaced by $N_c'$.
The PQ symmetry in this sector is also broken spontaneously at the dynamical scale of $SU(N_c')$.

In this model, the anomalous couplings of the axion candidates are given by
\begin{eqnarray}
\label{eq:comp}
{\cal L} = \frac{g_s^2}{32\pi^2} 
\left(\frac{N_c\,\tilde a}{ f_a} + \frac{N_c'\,\tilde b}{ f_b}\right)
G^a\tilde{G}^a \ .
\end{eqnarray}
Here, the decay constants are taken so that the domains of $\tilde{a}/f_a = [0,2\pi)$
and $\tilde{b}/f_b = [0,2\pi)$ coincide with the domains of the axial components of the
quark bilinears, $Q\bar{Q}$ and $Q'\bar{Q}'$, respectively.
From Eq.\,(\ref{eq:comp}),  we find an anomaly free combination is given by taking
\begin{eqnarray}
\label{eq:gPQcomposite}
q'/q = - N_c/N_c' ,
\end{eqnarray}
with which we identify the $U(1)_{gPQ}$ gauge symmetry in our general prescription.
The anomalous $U(1)_{aPQ}$ symmetry is, on the other hand, given by 
Eq.\,(\ref{eq:aPQ}).
The axion domain wall 
number corresponds to the greatest common devisor of $N_c$ and $N_c'$.

Under the $U(1)_{gPQ}$ symmetry, there are explicit breaking terms of the $U(1)_{aPQ}$ symmetry,
\begin{eqnarray}
{\cal L} \sim \frac{1}{M_{\rm PL}^{3|q|+3|q'|-4}}(Q\bar{Q})^{|q'|}(Q'\bar{Q}')^{|q|}\ .
\end{eqnarray}
These operators does not spoil the PQ solution for $3(|q|+|q'|) > 10$.
Thus, for example, a model with $N_c = 2$ and $N_c' = 5$ provides the 
origin of the anomalous PQ symmetry for the successful PQ mechanism.

For $q = 3 k$ or $q' = 3 k'$($k, k' \in {\mathbb Z}\backslash\{0\}$), there are additional lower dimensional operators which break the $U(1)_{aPQ}$ symmetry,
\begin{eqnarray}
{\cal L} \sim \frac{1}{M_{\rm PL}^{|q|+3|q'|-4}}(Q\bar{Q})^{|q'|}(q'\bar{q}')^{|q|/3}\ ,
\end{eqnarray}
or 
\begin{eqnarray}
{\cal L} \sim \frac{1}{M_{\rm PL}^{3|q|+|q'|-4}}(q\bar{q})^{|q'|/3}(Q'\bar{Q}')^{|q|}\ ,
\end{eqnarray}
Those operators are harmless for  $|q|+3|q'| > 10$ or $3|q| + |q'|> 10$,
which can be satisfied for $N_{c} = 3$ and $N_c' = 4$ for example.
\vspace{-.4cm}

\section{Discussions}
\vspace{-.3cm}
In this paper, we made  an attempt to explain an origin of the anomalous global PQ symmetry.
In our prescription, the anomalous global PQ symmetry originates from the gauged $U(1)$ symmetry
where the PQ symmetry is virtually embedded in a gauged $U(1)$ symmetry. 
Due to its simplicity, this mechanism can be implemented in many conventional models with the PQ symmetry.

In this prescription, the anomalous PQ symmetry appears as an approximate symmetry.
Thus, it is expected that the PQ symmetry is broken not only by the QCD anomaly 
but by some very higher dimensional operators to some extent.
Thus, the effective $\theta$-angle at the vacuum of the axion field is expected to be non-vanishing completely,
though its numerical value highly depends on models.

As we have seen, our prescription allows models with either $N =1$ or $N> 1$.
For $N > 1$, the axion potential generated by the non-perturbative QCD effects has 
a ${\mathbb Z}_{N} (\subset U(1)_{aPQ})$ symmetry.
When the ${\mathbb Z}_{N}$ symmetry is an exact symmetry, models with $N >1$ causes 
a serious domain wall problem if  spontaneous breaking of the PQ symmetry takes place 
after inflation
\footnote{In our prescription, however, ${\mathbb Z}_{N}$ is not an exact symmetry, and hence,
the domain wall problem associated for $N > 1$ might be avoidable by the effects 
of the explicit $U(1)_{aPQ}$ symmetry breaking~\cite{Hiramatsu:2010yn}.}. On top of the above arguments, there can also be a serious domain wall problem even for $N$ = 1~\cite{Barr:1986hs}.

A trivial solution to the domain wall problems is to assume that the PQ symmetry breaking takes place
before the end of inflation.
In this case, the Hubble constant during inflation is limited from above 
to avoid the so-called isocurvature problem~\cite{Axenides:1983hj,Seckel:1985tj,Linde:1985yf,Linde:1990yj,Turner:1990uz,Lyth:1991ub}.

\vspace{-.4cm}

\section*{Acknowledgements}
\vspace{-.3cm}
This work is supported in part by Grants-in-Aid for Scientific Research from the Ministry of Education, Culture, Sports, Science, and Technology (MEXT) KAKENHI, 
Japan, No.\,25105011 and No.\,15H05889 (M. I.) as well as No.\,26104009 (T. T. Y.); Grant-in-Aid No.\,26287039 (M. I. and T. T. Y.) and  No.\,16H02176 (T. T. Y.) 
 from the Japan Society for the Promotion of Science (JSPS) KAKENHI; and by the World Premier International Research Center Initiative (WPI), MEXT, Japan (M. I., and T. T. Y.).
 The work of H.F. is supported in part by a Research Fellowship for
 Young Scientists from the Japan Society for the Promotion of Science (JSPS).

\bibliography{../papers}
\nocite{Fayet:1977yc}

\end{document}